\documentclass[useAMS,usenatbib]{mn2e}
\usepackage{graphicx}
\usepackage{epsfig}
\usepackage{amssymb}
\usepackage{lscape}
\usepackage{ulem}
\usepackage{txfonts}
\usepackage{natbib}
\usepackage{url}

\def\kms{km~s$^{-1}$}
\def\hkpc{$h_{70}^{-1}$ kpc}

\def\dv{$\Delta$V}

\def\rp{$r_p$}

\title[\textit{WISE} AGN] {Galaxy pairs in the Sloan 
Digital Sky Survey - IX:  Merger-induced AGN activity as traced by the
\textit{Wide-field Infrared Survey Explorer}.}

\author[Satyapal et al.] {Shobita Satyapal$^1$,
Sara L. Ellison$^2$,
William McAlpine$^1$,
Ryan C. Hickox$^3$,
\newauthor David R. Patton$^4$,
J. Trevor Mendel$^5$.\\
$^1$ George Mason University, Department of Physics, Astronomy, \& Computational Sciences, MS 3F3, 4400 University Drive, Fairfax, VA 22030, USA\\ 
$^2$ Department of Physics \& Astronomy, University
of Victoria, Finnerty Road, Victoria, British Columbia, V8P 1A1,
Canada.\\
$^3$ Department of Physics and Astronomy, Dartmouth College, 6127 Wilder Laboratory, Hanover, NH 03755, USA \\
$^4$ Department of Physics \& Astronomy, Trent University,
1600 West Bank Drive, Peterborough, Ontario, K9J 7B8, Canada.\\
$^5$ Max-Planck-Institut fur Extraterrestrische Physik, 
Giessenbachstrasse, D-85748 Garching, Germany.
}
\begin{document}

\maketitle

\begin{abstract}

Interactions between galaxies are predicted to cause gas inflows that
can potentially trigger nuclear activity. Since the inflowing material
can obscure the central regions of interacting galaxies, a potential
limitation of previous optical studies is that obscured Active
Galactic Nuclei (AGNs) can be missed at various stages along the
merger sequence.  We present the first large mid-infrared study of
AGNs in mergers and galaxy pairs, in order to quantify the incidence of obscured AGNs
triggered by interactions.  The sample consists of galaxy pairs and
post-mergers drawn from the Sloan Digital Sky Survey that are matched
to detections by the Wide Field Infrared Sky Explorer
(\textit{WISE}). We find that the fraction of AGN in the pairs,
relative to a mass-, redshift- and environment-matched control sample,
increases as a function of decreasing projected separation.  This
enhancement is most dramatic in the post-merger sample, where we find
a factor of 10-20 excess in the AGN fraction compared with the control.
Although this trend is in qualitative agreement with results based on
optical AGN selection, the mid-infrared selected AGN excess increases much more
dramatically in the post-mergers than is seen for optical AGN. Our
results suggest that energetically dominant optically obscured AGNs become more prevalent in
the most advanced mergers, consistent with theoretical predictions.
\end{abstract}


\section{Introduction}

Based on the current cold dark matter cosmological framework, it is
now well-established that galaxy interactions are ubiquitous and that
they play a pivotal role in the formation and evolution of
galaxies. From both a theoretical and observational perspective,
galaxy interactions are undoubtedly responsible for enhanced nuclear
star formation
\citep[e.g.,][]{mihos1996,larson1978,sanders1996,kennicutt1987,per06,woods2006,woods2007,dim07,dim08,ellison2008,cox08,smith2010,patton2011,liu12,scudder2012,patton2013},
and the formation of spheroids
\citep[e.g.,][]{toomre1977,lake1986,shier1998,rothberg2006}. A natural
assumption from the tight correlation between central black hole mass
and bulge velocity dispersion \citep[e.g.,][]{gebhardt2000} is that in
addition to bulge growth, interactions trigger accretion onto a
central supermassive black hole. However, despite over three decades
of extensive research, it is still a topic of debate whether or not
there is observational evidence for a causal connection between
mergers and Active Galactic Nuclei (AGN), and, if so, how this
connection depends on merger and host galaxy parameters. A number of
studies have found evidence for mergers in luminous quasar hosts
\citep[e.g.,][]{canalizo2001,benn08,urru08,for09,ramos2011,bess12,urru12},
where the link to interactions is fairly well accepted. On the other
hand, the connection to the less luminous AGN population remains
controversial.  In particular, studies that have looked for an excess
of tidal features or distorted morphologies in AGN versus non-AGN galaxies, have found no
statistical difference
both at low \citep[e.g.,][]{gab09,rei09,cisternas2011,koc11,bohm12}, intermediate(0.5$<$z$<$0.8) \citep[e.g.,][]{villforth2014} and higher (z$>$1) redshifts \citep[e.g.,][]{karouzos2014,fan2014} .  Conversely,
studies of close pairs have found enhanced fractions of AGN (or
accretion rates), which supports a link between mergers and nuclear
activity
\citep[e.g.,][]{alo07,woods2007,koss2010,ellison2011,sil11,koss2012,liu12,ellison2013a,sbaf13}.
This discrepancy may be due in part to the low surface brightness of
tidal features and the time during the interaction at which they are
expected to be visible \citep[e.g.][]{lotz08}. If the luminosity of the AGN is variable over a wide dynamic range on timescales shorter than the lifetime of merger signatures, any observed trends of merger fraction as a function of AGN luminosity will be weak, while the incidence of AGNs in merging galaxies will be still higher than in isolated systems \citep{hickox2014}.
 Furthermore, since tidal features can be faint and appear only
in the gas instead of the stars \citep[e.g.][]{kuo2008}, the
sensitivity \citep[e.g.][]{canalizo2001,ramos2011} and the wavelength
\citep[e.g.,][]{hancock2007,boselli2005} of the observations may play a
role in identifying merger signatures in AGN hosts.  However, an alternative
way of reconciling the apparently conflicting results is if mergers 
\textit{can} trigger AGN, but the majority of AGN are not produced
through an interaction \citep[e.g.][]{db12}.  

It is now well-known  that observations in only one waveband cannot provide a complete census of AGNs in galaxy samples due to obscuration of the central source or contamination of the observed emission by the host galaxy \citep[e.g.][]{satyapal2008,goulding2009,hickox2009,donley2010,juneau2013}. Although AGN excesses in samples of galaxy pairs have been
found at optical, radio and X-ray wavelengths
\citep[e.g.,][]{woods2007,koss2010,sil11,koss2012,liu12,sbaf13}, a
direct comparison of these selection techniques has not been
previously performed, and we have little understanding of what the
complete census of merger-induced AGN might be.  In our previous work
on galaxy pairs and post-mergers in the Sloan Digital Sky Survey, we
have used optical emission line diagnostics to identify an
enhanced AGN fraction relative to a control that increases with decreasing pair separations \citep{ellison2011} and peaks post-coalescence
\citep{ellison2013a}.  However, since the centres
of interacting galaxies may be more obscured than isolated galaxies, a
potential limitation of this and previous optical studies, is that
obscured AGNs can be missed at various stages along the merger
sequence \citep{goulding2012}.  In such cases, mid-infrared observations are a powerful
tool for finding optically obscured AGNs. While there have been a
number of mid-infrared studies of interacting galaxies, virtually all
past studies have employed small samples of galaxies and/or have
targeted the most advanced stage mergers
\citep[e.g.,][]{genzel98,veilleux2009,armus07,armus09,farrah07,petric11}. The
all-sky survey carried out by the {\it Wide-field Infrared Survey
Explorer (WISE)} \citep{wright2010} has opened up a new window in the search for
optically hidden AGNs in a large number of galaxies. This is because
hot dust surrounding AGNs produces a strong mid-infrared continuum and
infrared spectral energy distribution (SED) that is clearly
distinguishable from star forming galaxies in both obscured and
unobscured AGNs \citep[e.g.][] {lacy2004,stern2005,donley2007,stern2012}.  The \textit{WISE} survey
enables a more statistically significant study of the optically obscured AGN
population in interacting galaxies.

The goal of this paper is to complement our previous optical AGN study
of SDSS galaxy pairs with a measurement of the incidence of obscured
AGN, using mid-infrared colour selection with \textit{WISE}.  This is the first large mid-infrared study of galaxy pairs. In Section
\ref{sample_sec} we describe the selection of our samples of galaxy
pairs, post-mergers, and their matched controls. In Section \ref{WISE_sec}, we discuss our
\textit{WISE} AGN classification criteria, followed in Section
\ref{IRAC_sec} by a discussion of the fidelity of our {\it WISE}
photometry for close pairs.   in Section
\ref{results_sec},  we determine the
mid-infrared colour-selected AGN fraction in the merger samples using
\textit{WISE} compared to the control sample. In Section \ref{other_sec} we discuss other causes of red
\textit{WISE} colours, followed by a summary of our results in Section
\ref{summary_sec}.  Throughout the manuscript we adopt a cosmology with $H_0$ = 70 \kms $,\Omega_M=0.3$, and $\Omega_{\Lambda}$=0.7.

\section{Sample Selection}
\label{sample_sec}

The galaxy merger sample is based on a combination of close spectroscopic
galaxy pairs and visually classified post-mergers, which represent the
early and late stages of galaxy interactions, respectively.  The
sample is described in detail in \citet{ellison2013a}.  In brief, the
pairs sample is constructed from the SDSS DR7 Main Galaxy Sample (14.0
$\le m_r \le$ 17.77) with a redshift range $0.01 \le z \le 0.2$ and spectroscopically classified as a galaxy (specclass=2).  We require projected separations of \rp\ $\le$ 80
\hkpc, relative velocities  of \dv $\le$ 300 \kms\ and ratios of stellar
mass taken from \citet{mendel2014} of 0.25 $\le$ M$_1$/M$_2$ $\le$ 4.
Culling of wide separation pairs accounts for fibre collisions
\citep{ellison2008,patton2008}.  We note that not all paired galaxies show visible signs of interacting, and many may never merge. The sample of visually classified
post-mergers is initially drawn from the Galaxy Zoo \citep{lintott2008}
catalogue presented by \citet{darg2010a}, with further visual
inspection and refinement by \citet{ellison2013a}.  For convenience, we
will refer collectively to the pairs and the post-mergers as the
`merger' sample.

The merger sample is matched to the public final all-sky \textit{WISE} source
catalogue\footnote{\url{http://wise2.ipac.caltech.edu/docs/release/allsky/}}, where a galaxy is considered `matched' if the
positions agree to within 6 arcseconds (the resolution of
\textit{WISE})\footnote{The conclusions in this paper are unchanged if
a more strict matching criterion of 2 arcseconds is used, although the
statistics are slightly poorer.}.  For the majority of this paper, we
will use the 3.4 $\mu$m and 4.6$\mu$m bands (W1 and W2 respectively)
to classify AGN.  We therefore additionally require that WISE matches
to SDSS galaxies are detected at better than 5$\sigma$ in each of
these two bands.  After \textit{WISE} matching, we are left with 80
post-mergers and 5026 galaxies in the pairs sample.

In order to compare the merger sample to a control sample, we follow
the procedure described in \citet{ellison2013a}.  In brief, every
galaxy that is matched to a \textit{WISE} source and has no spectroscopic
companion within 80 \hkpc\ and a relative velocity $\Delta V$
within 10,000 \kms, and has a Galaxy Zoo merger vote fraction = 0, is
considered as part of a control `pool'.  There are 204, 596 galaxies in
the control pool.  For a given galaxy in the merger sample, we compile
a set of controls that are matched in stellar mass, redshift and local
environment.  This latter parameter is defined as

\begin{equation}
\Sigma_n = \frac{n}{\pi d_n^2},
\end{equation}

where $d_n$ is the projected distance in Mpc to the $n^{th}$ nearest
neighbour within $\pm$1000 \kms.  Normalized densities, $\delta_n$,
are computed relative to the median $\Sigma_n$ within a redshift slice
$\pm$ 0.01.  Following our previous work, we adopt $n=5$.  The
tolerances for matches are 0.1 dex in stellar mass, 0.005 in redshift
and 0.1 dex in $\delta_5$.  If less than 5 matches are found for a
given merger galaxy, the tolerance is grown by 0.005 in
redshift, 0.1 dex in stellar mass and 0.1 dex in normalized local
density until the required number of matches is achieved.  In practice,
several hundred control galaxies are typically matched to each merging galaxy,
without the need for extending the baseline tolerances.

\section{Selection of AGNs by \textit{WISE}}\label{WISE_sec}

Extensive efforts over the past decade have demonstrated the power and
reliability of mid-infrared observations in discovering optically
hidden AGNs \citep[e.g.,][]{lacy2004,stern2005, donley2007,hickox2007,donley2008, eckart2010,stern2012, mateos2013}. This is because hot dust surrounding AGNs produces a
strong mid-infrared continuum and infrared SED 
that is clearly distinguishable from normal
star-forming galaxies for both obscured and unobscured AGNs in
galaxies where the emission from the AGN dominates over the host
galaxy emission \citep[e.g.][]{donley2007,stern2012,mateos2013}.  In particular, at low
redshift, the $W1$ (3.4 $\mu$m) - $W2$ (4.6 $\mu$m) colour of galaxies
dominated by AGNs is considerably redder than that of inactive
galaxies \citep[see Figures 1 in][]{stern2012,assef2013}.  At higher
redshifts ( z $>$ 1.5) the host galaxy becomes red across these bands
but becomes undetectable by \textit{WISE}.

There are several \textit{WISE} colour diagnostics that have been
employed in the literature to select AGNs.  Based on the
\textit{Spitzer-WISE} COSMOS data, \citet{stern2012} show that a
mid-infrared colour cut of $W1-W2 >0.8$ robustly identifies AGNs
previously identified by \textit{Spitzer} with a reliability of
95\%.  However, at redshifts below 0.2, even the most
extreme star forming templates from \citet{assef2013}, have $W1-W2$
colour well below 0.5. In this work, we adopt a robust colour cut of $W1-W2 > 0.8$ to identify AGNs. Given that the redshifts of the galaxies in our
pairs sample are less than 0.2, we also discuss the results with a more inclusive colour cut of $W1-W2>0.5$ and demonstrate that our results are qualitatively unchanged.

We emphasize that while mid-IR colour selection is sensitive to the optically
obscured AGN population and can also select unobscured AGNs, it is
sensitive only to AGNs that dominate over the host galaxy emission that are efficiently accreting \citep[e.g.,][]{gurkan2014}.
Based on the templates from \citet{assef2010}, the $W1-W2$ colour
drops below 0.8 when the host galaxy emission exceeds 50\% of the
total galaxy light.  When the host galaxy contamination of the
mid-infrared emission is large, the $W1-W2 > 0.8$ colour cut adopted
here will not select the AGN.  This effect is born out observationally
in recent studies that show that the fraction of optically and X-ray
identified AGNs identified by {\it WISE} is highest for the most
luminous AGNs (based on X-ray or [OIII] luminosity) and drops
considerably with decreasing AGN luminosity \citep[e.g.,][]{mateos2013,rovilos2013}

\section{The fidelity of \textit{WISE} colours in close pairs}\label{IRAC_sec}

The analysis presented in this paper relies on the measurement of
colours in close pairs of galaxies.  As we have previously found
for the SDSS photometry, contamination from the companion becomes
problematic at small angular separations \citep{simard2011}.  Since 
\textit{WISE} has an even lower spatial resolution that the SDSS
(6.1 and 6.4 arcsec in the W1 and W2 bands used in this work),
we might expect blending to also affect the \textit{WISE} photomerty.

We investigate the possible effects of contamination by
utilizing {\it Spitzer} Infrared Array Camera (IRAC) observations that
cover very similar wavelength ranges to the W1 and W2 bands, but whose spatial
resolution is 1.7 arcsec in the bands considered here. We
use data from the {\it Spitzer} Deep Wide-Field Survey (SDWFS;
\citet{ashby2009}), who performed deep observations with IRAC over 8
deg$^2$ in the Bo\"{o}tes multiwavelength survey area. We select
bright SDWFS sources with $[3.6]<16$, $\approx$1 magnitude fainter
than the typical {\it WISE} limit used in this study, and therefore
the flux above which contamination may be significant. We do not
include IRAC sources near bright stars and other areas of bad IRAC
photometry \citep[e.g.,][]{hickox2011}. We match the {\it WISE}
All-Sky Survey source positions to these bright IRAC sources, with a
matching radius of 2\arcsec, and compare the {\it WISE} ${\mathrm
W1}-\mathrm{W2}$ colours to the $[3.6]-[4.5]$ colours of the matched
IRAC sources. To accurately compare the colours, we correct for a known colour dependence in the difference between {\it WISE} and IRAC photometry \footnote{\url{http://wise2.ipac.caltech.edu/docs/release/allsky/expsup/}}. For the sources in our sample, red objects ($[3.6]-[4.5] > 0.3$) have colours such that $W1 - [3.6] \approx 0.4([3.6]-[4.5])$ and $W2 - [4.5] \approx 0.05([3.6]-[4.5])$, and we apply these corresponding offsets to the observed IRAC magnitudes.We further calculate the distance from each matched IRAC
source to the closest other bright source in the SDWFS catalogue; for
closer IRAC pairs (particularly within $\sim$10\arcsec) we may expect
stronger contamination of the observed {\it WISE} flux from the
companion source.

\begin{figure}
\centerline{\rotatebox{0}{\resizebox{9cm}{!}
{\includegraphics{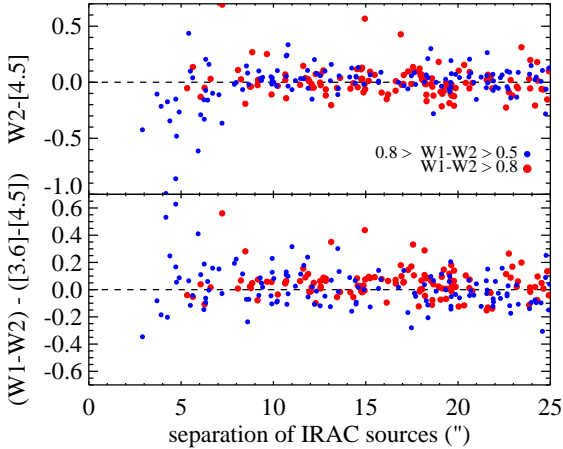}}}}
\caption{\label{wiseirac} Top panel: Difference between the {\it WISE} W2 magnitude and the IRAC [4.5] magnitude for AGN candidates as a function of distance from the matched IRAC source to the nearest bright IRAC source. Bottom panel: Difference between the {\it WISE} $W1-W2$  and the IRAC $[3.6]-[4.5]$ colours for AGN candidates as a function of IRAC separation. Galaxies with $\mathrm{W1-W2}>0.8$ are shown in red, while galaxies with $\mathrm{W1-W2}>0.5$ are shown in blue; see Section \ref{WISE_sec}. }
\end{figure}

The contamination in the \textit{WISE} photometry is demonstrated in
the top panel of Figure \ref{wiseirac}, which shows the difference
between the {\it WISE} W2 and IRAC [4.5] magnitudes for AGN candidates as a function of
distance from the matched IRAC source to the nearest bright
($[3.6]>16$) IRAC source. AGN candidates with $W1-W2  > 0.8$ and  $0.5 < W1-W2 < 0.8$ are marked with red and blue symbols, respectively. For large IRAC pair separations the average
${\mathrm W2}-[4.5]\approx 0$ .  For separations below
$\sim$10\arcsec, the {\it WISE} flux becomes increasingly brighter
than the flux of the matched IRAC source, by up to $\sim$0.5-1 mag,
indicating that blending of the two IRAC sources contaminates the {\it
WISE} fluxes. 

However, if we examine the difference in {\it WISE} and
IRAC {\it colours} as a function of IRAC separation (bottom panel of
Figure \ref{wiseirac}), there is no significant change in the average
colour difference even to small separations $\lesssim5$\arcsec, and
only a small increase in the number of outliers; for 366 IRAC sources in pairs with separation $< 10\arcsec$ , the dispersion in the colour differences ($\sigma=0.22$) is only slightly larger than that for all separations (0.17). Of these IRAC sources with close companions, 15 have WISE colours consistent with AGN ($W1-W2 > 0.8$; see Section \ref{WISE_sec}), and in 13 cases, the matched IRAC source would have been identified as an AGN based on the \citep{stern2005} criterion. We therefore conclude that for close pairs of mid-IR sources, contamination of WISE photometry due to unresolved mid-IR companions has a small but relatively unimportant effect on the selection of AGNs based on observed WISE colours.

In addition to the tests described above, two other observational
factors mitigate the susceptibility of our conclusions to
contamination.  First, whilst previous studies of the optical colours
of close pairs have shown an effect in the mean values
\citep{patton2011}, we will quantify the \textit{fraction} of galaxies
with colours above a certain threshold.  The small magnitude of colour
changes is unlikely to shift a large number of galaxies from the star-forming
to AGN class; there is at least 0.2 mags difference between these
two classes in W1 -- W2 colour space.  Second, even if we exclude
pairs with angular separations $<$ 5 arcsec, our basic result remains
unchanged.  This is largely thanks to the inclusion of post-mergers
in our sample, giving us a data point that probes the final stage of
the interaction without being susceptible to  contamination from a
near neighbour.

\section{\textit{WISE} AGN fraction in the merger sample}\label{results_sec}

\begin{figure}
\centerline{\rotatebox{0}{\resizebox{9cm}{!}
{\includegraphics{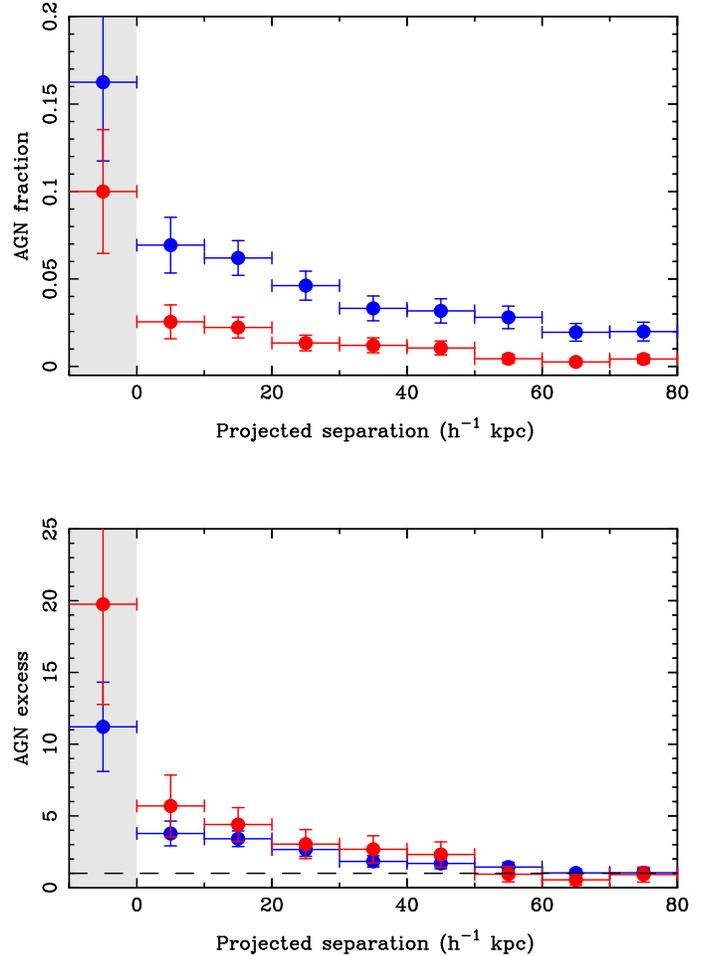}}}}
\caption{\label{agn_thresh} \textit{WISE} AGN fraction (top panel) and AGN excess (bottom panel)  by projected pair separation for the pairs sample and post-merger galaxy sample at the $W1-W2 >0.8$ threshold (red points) and at the $W1-W2 >0.5$ threshold (blue points). The grey shaded region corresponds to the post-merger sample.}
\end{figure}

 Applying a \textit{WISE} colour criteria of $W1-W2>0.8$ to identify
 {\it WISE}-AGNs in our merger samples, we find that 52 of 5026 (1\%) paired galaxies
 and 7 of 80 (9\%)post-mergers are {\it WISE}-AGNs, compared with 0.5\%  the corresponding matched controls.  In Figure \ref{agn_thresh}, we plot the AGN fraction and the AGN excess, which is the fraction of AGNs
 in the pairs sample relative to their controls, as a function of pair
 separation .
 There is a steady increase in the AGN excess with decreasing pair
 separations for galaxies at separations less than 50 \hkpc\, with a
 dramatic increase seen in the post-merger sample (shaded region in
 Figure \ref{agn_thresh}).  Since the redshift range of our sample is low, and the colour cut of $W1-W2 >0.8$ only selects the most energetically dominant AGNs (see Section \ref{WISE_sec}), we also repeat the calculation for the less stringent colour cut of $W1-W2 >0.5$ (blue points in figure). As can be seen, the AGN excess is replicated using both colour cuts, with a moderately higher excess seen with the more stringent colour cut (red points in figure).  Indeed, in the post-merger
 bin, AGNs with $W1-W2>0.8$ are 20 times more frequent than their
 matched control sample, compared to an excess of $\sim$ 11 obtained
 with the $W1-W2>0.5$ AGNs. Since the mid-infrared colour of galaxies
 increases as the AGN contribution to the galaxy light increases, the
 larger AGN excess seen with the more stringent colour cut, most
 notably in the post-merger sample, is likely due to an increase in
 the AGN contribution to the total galaxy light.


\subsection{Comparison with Optical AGN}



\begin{figure}
\centerline{\rotatebox{0}{\resizebox{7cm}{!}
{\includegraphics{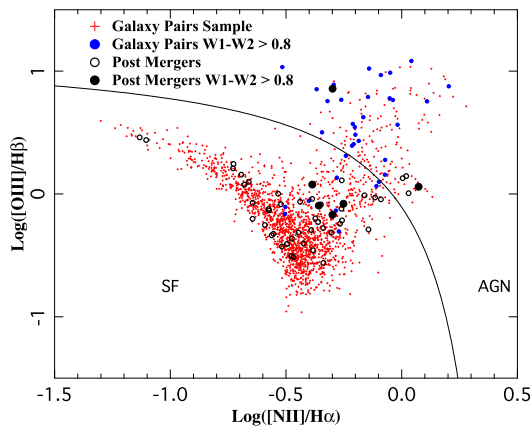}}}}
\caption{\label{optical_bpt} BPT diagram for those galaxies in the pairs and post-merger samples with signal to noise $>5~\sigma$ in all four emission lines.  The solid line corresponds to the AGN demarcation from \citet{kewley2001}.}
\end{figure}

We compared the {\it WISE} AGN selection with optical emission line
selection using the [NII]/H$\alpha$ versus [OIII]/H$\beta$, line
ratios using the widely-adopted BPT diagram (Baldwin, Phillips \&
Terlevich 1981). In Figure \ref{optical_bpt}, we plot the optical BPT
diagram for the pairs and post-merger samples.  Only those galaxies
with all four emission lines detected with a signal to noise $>5\sigma$
are included. Using the classification scheme from \citet{kewley2001},
there are 200 optically identified AGNs in the pairs sample, of which
175 are not identified as AGN using the $W1-W2 >0.8$ colour cut. This
number reduces to 138 with the less stringent $W1-W2>0.5$ colour
cut.

\begin{figure}
\centerline{\rotatebox{0}{\resizebox{9.5cm}{!}
{\includegraphics{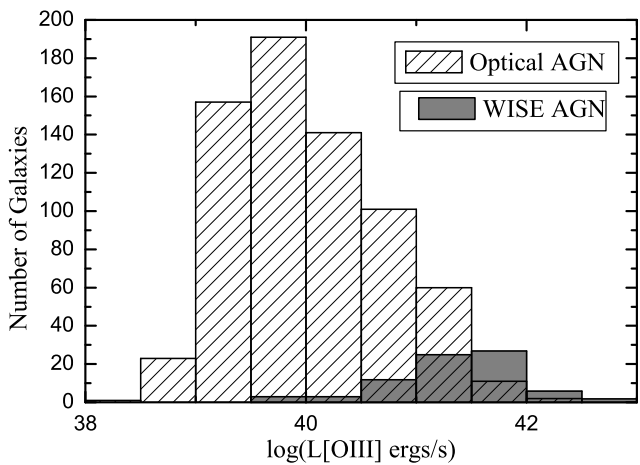}}}}
\caption{\label{oiii} [OIII] luminosities of optically selected AGN in the pairs sample that are \textit{WISE} AGN ($W1-W2>0.8$) and of optically selected AGN that are not \textit{WISE} AGN}
\end{figure}

It is clear that \textit{WISE} selects a different population of AGNs
than does the optical emission line diagnostics.  As pointed in
Section 4, mid-infrared colour selection identifies AGNs when the AGN
dominates the host galaxy emission.  This is reinforced by several
studies that have shown that the fraction of optically selected AGNs
identified as AGNs by mid-IR colour selection increases with AGN
luminosity  \citep[e.g.,][]{rovilos2013,mateos2013,yan2013} .  We find a similar trend with \textit{WISE} AGN fraction
and AGN luminosity for the optically classified AGNs in our pairs
sample. For the optical AGN in our pairs sample we used the [OIII]
$\lambda$5007 emission line as an indicator of the luminosity of the
AGN. The [OIII] $\lambda$5007 line is found to be well correlated with
the AGN bolometric luminosity in optically classified AGNs
\citep{heckman2004}. As seen in Figure \ref{oiii} the optical AGNs
that are \textit{WISE} AGNs tend to have higher [OIII] luminosities
than optical AGN that are not \textit{WISE} AGN. The most luminous AGNs in the sample are all identified as WISE AGNs. For [OIII]
luminosities greater than 10$^{42}$ ergs s$^{-1}$ we find that 60\% of
the optical AGN are identified as \textit{WISE} AGN. For luminosities
between 10$^{41}$ and 10$^{42}$ ergs s$^{-1}$, the \textit{WISE} AGN
fraction drops to 16\% and to 1\% for [OIII] luminosities less than
10$^{41}$ ergs s$^{-1}$. Based on this analysis, {\it WISE} colour
selection identifies the more powerful AGNs in our samples.

\begin{figure}
\centerline{\rotatebox{270}{\resizebox{7cm}{!}
{\includegraphics{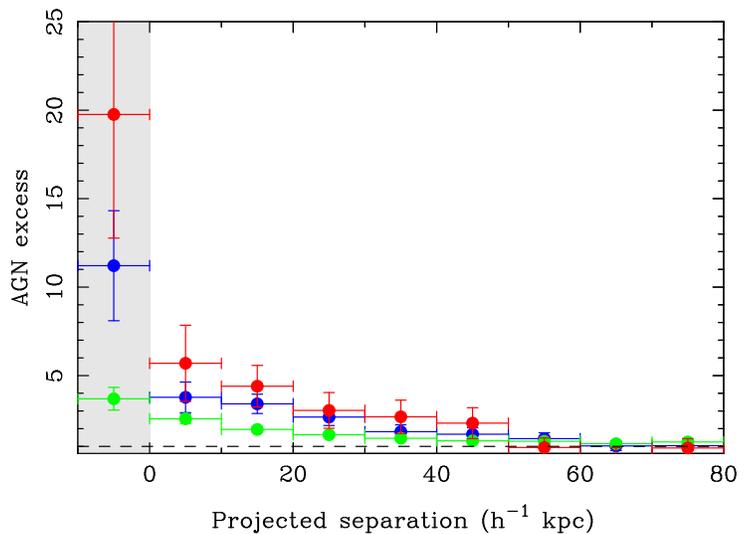}}}}
\caption{\label{excess_opt} The excess of merger AGN classified by \textit{WISE} using our  $W1-W2>0.8$ colour cut (red) and $W1-W2>0.5$ colour cut (blue) compared
to the excess determined from optical emission lines (green) as presented in \citet{ellison2013a}.}
\end{figure}

\begin{figure}
\centerline{\rotatebox{270}{\resizebox{6cm}{!}
{\includegraphics{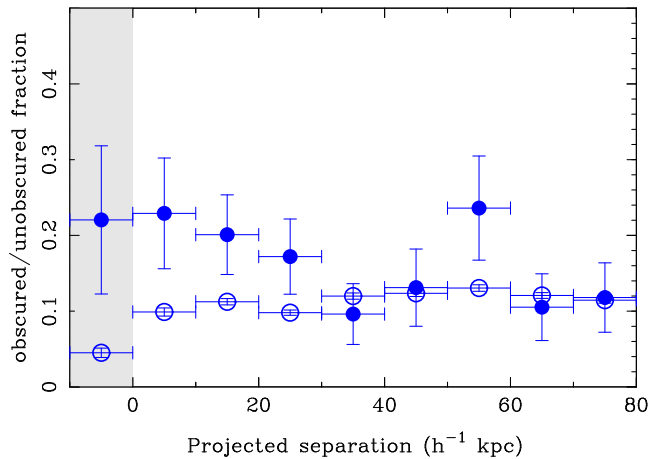}}}}
\caption{\label{dust_excess} The ratio of the fraction of galaxies classified as AGN only by \textit{WISE} to the fraction of optical AGN as a function of pair separation for the pairs (solid symbols) compared with the controls (open symbols).}
\end{figure}

In Figure \ref{excess_opt} we compare the excess of merger AGN classified by \textit{WISE},
to the excess determined from optical emission lines, as presented by
\citet{ellison2013a}.  The optical classification uses the diagnostic
of \citet{stas2006} which is sensitive to even modest
contributions from an AGN.  It should be noted that, as with any criterion
for AGN classification, the exact AGN fraction depends sensitively on
diagnostic choice.  For optical classifications, the choice of emission
line S/N threshold also plays a role \citep{ellison2011}.  Comparing
the exact fractions of AGN between the \textit{WISE} and optical
classifications is therefore not instructive.  However, a comparison
of the AGN excess in the optical and \textit{WISE} samples as a function of
projected separation can tell us about the evolving properties of
merger-induced AGN as a function of interaction stage.

As can be seen from Figure \ref{excess_opt}, the trend of increasing
{\it WISE}-AGN excess with decreasing pair separation, peaking in the
post-merger sample, is in {\it qualitative} agreement with results
based on optical AGN selection. However, the AGN excess based on
infrared colour selection is significantly larger than the AGN excess
based on optical spectroscopic diagnostics for the galaxy pairs at the
smallest pair separations, with the most dramatic discrepancy seen in
the post mergers. This is most extreme for the more stringent $W1-W2>0.8$ colour cut, which selects the most energetically dominant AGNs. This plot implies that the AGNs are more energetically dominant with decreasing pair separation.  In Figure \ref{dust_excess}, we plot the ratio of the fraction of galaxies that are identified as AGNs only with WISE to the fraction of galaxies that are optical AGNs in the pairs (solid symbols) compared with the matched controls (open symbols) as a function of pair separation.  Since we have demonstrated that the observed trends are reproduced using both colour cuts discussed in this paper, for this figure we  adopt a $W1-W2>0.5$ colour cut to improve the signal to noise of the plot. Note that {\it WISE} identifies AGNs only if they are energetically dominant regardless of whether or not they are obscured in the optical \citep{stern2012}. However, if a galaxy is selected as an AGN with {\it WISE} but optically unidentified as an AGN, its broad and narrow line regions are optically obscured.   Figures  \ref{excess_opt} and \ref{dust_excess} therefore together imply that the AGNs are : 1) more energetically dominant and 2) more heavily obscured over large scales with decreasing pair separation.  Theory predicts that as a merger progresses,
gravitational instabilities cause large radial gas inflows toward the
nuclear regions.  This inflowing material can potentially obscure both the narrow and broad line regions of AGNs, causing any putative AGN to be unidentified at optical wavelengths. These numerical simulations show that multiple gas
inflow epochs along the merger sequence occur, with enhanced nuclear
accretion in the most advanced mergers, just prior to coalescence
\citep[e.g.,][]{mihos96,cox06,barnes96,cox08}. Our results are
consistent with these theoretical predictions, suggesting that AGNs
are both more obscured and more energetically dominant in the most advanced
mergers.

\begin{figure}
\centerline{\rotatebox{0}{\resizebox{9cm}{!}
{\includegraphics{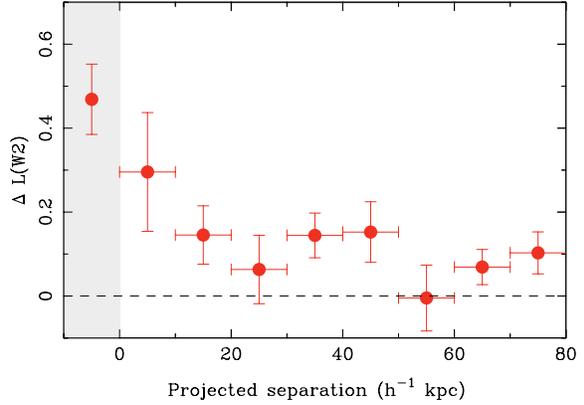}}}}
\caption{\label{w2} The enhancement in the W2 band luminosity for close pairs of \textit{WISE} AGN
galaxies relative to their control sample. The point in the grey shaded box
shows the enhancement for galaxies in the post-merger sample.}
\end{figure}

To investigate further the effect of a merger on the AGN luminosity we
determined the excess W2 band luminosities of our pairs and
post-merger sample \textit{WISE}-AGN compared to a matched control
sample of \textit{WISE}-AGN.  In galaxies dominated by an AGN, the W2
luminosity is well-correlated with the AGN bolometric luminosity
\citep{richards2006}.  In order to calculate the AGN contribution
in mergers, relative to that in isolated AGN, we repeated the
control matching process, but now only match control galaxies that
are also classified as AGN, according to our \textit{WISE} colour
criterion.  We can now compare the properties of merger-induced
AGN with control AGN.  Using the luminosity in the W2 band \footnote{\url{http://wise2.ipac.caltech.edu/docs/release/allsky/expsup/}}
, we compute the offset between the merger and control AGN
luminosities as

\begin{equation}
\Delta L(W2) = \log (L(W2)_{merger}/L(W2)_{control})
\end{equation}

where $L(W2)_{control}$ is the median luminosity of all controls
matched to a given merging galaxy.  Figure \ref{w2} shows the
enhancement in the W2 band luminosity of the pairs and post-merger
galaxies.  In order to improve the signal to noise of the plot, we have adopted the $W1-W2 > 0.5$ colour cut. The AGNs in galaxy pairs show a significant enhancement of
W2 band luminosity out to 80 \hkpc, which is the widest separation in
our sample, with an enhancement of a factor of 2 for pairs separated
by less than 10 \hkpc\ .The post-mergers show a clear enhancement of
W2 luminosity by a factor of 3.  This is consistent with the results
from \citep{ellison2013a} which show a similar trend using the [OIII]
luminosity of optical AGN indicative of enhanced black hole accretion
rate for close pairs and post-mergers over isolated galaxies.

\section{Other Causes of Red \textit{WISE} Colours}\label{other_sec}

Ultra-luminous infrared galaxies (ULIRGs) are defined to have a total
infrared luminosity log L$_{\rm IR} > 12$ L$_{\odot}$, with their
lower luminosity cousins, the luminous infrared galaxies (LIRGs) a
factor of 10 less luminous.  ULIRGs may also produce red W1-W2 colours
\citep{wright2010} and therefore may be misclassified as an AGN,
although many ULIRGs may also host obscured AGNs \citep{veilleux2009}. We matched our pairs
sample to the 1 Jy sample, a flux-limited sample of ULIRGs from
\citet{kim1998},and we found only 1 match. ULIRGs are almost exclusively advanced mergers \citep{sanders1996, veilleux2009}. The low incidence of ULIRGs in our pairs sample is a consequence of the deficiency of very close pairs in the
sample. The results from this matching procedure are in good agreement
with \citet{ellison2013b}, who studied the fraction of LIRGs in the
close pairs sample.  Although the fraction of LIRGs was found to
increase with decreasing projected separation, and the fraction of
galaxies in pairs found to increase with increasing L$_{\rm IR}$, only
1 ULIRG was identified in the pairs sample.  A sub-sample of 74 ULIRGs
from \citet{veilleux2009} reveal that over half of the ULIRGs have
separations of less than 3 kpc and 86\% have separations less than
10kpc. In contrast only 3\% of our pairs sample have separations less
than 10kpc with none less than 3kpc. Therefore the fraction of
\textit{WISE} AGN in our pairs sample is not significantly affected by the presence
of ULIRGs with red W1-W2 colours.

\section{Summary}\label{summary_sec}

We have conducted a mid-infrared study aimed at finding obscured AGNs
using {\it WISE} matched to a large sample of galaxy pairs and post-mergers
selected from the SDSS.
This is the first mid-infrared investigation of a large sample of
galaxy pairs. Our main results can be summarized as follows:

\begin{enumerate}
\item{ We find a higher fraction of AGNs in galaxy pairs compared to a
carefully constructed control sample of isolated galaxies matched in
redshift, mass and local environment for pair separations less than 50 \hkpc\ .}

\item{ The excess in the AGN fraction over the matched control
increases with decreasing pair separation.  The excess is most
dramatic in the post-merger sample, where we find a factor of 10-20
excess in the AGN fraction compared to the control, depending on
the adopted colour threshold.}

\item{The trend of increasing infrared selected AGN fraction with
decreasing pair separation, peaking in the post-merger sample, is in
qualitative agreement with results based on AGN selection obtained
from optical emission line diagnostics. However, the AGN excess based
on infrared colour selection is significantly larger than the AGN
excess based on optical spectroscopic diagnostics for the galaxy pairs
at the smallest pair separations, with the most dramatic discrepancy
seen in the post mergers. Our results imply that AGNs are both more energetically dominant and obscured with decreasing pair separation, as expected based on theoretical predictions. }

\item{The AGNs in galaxy pairs show a significant enhancement of W2
band luminosity compared to their matched control out to at least 80
\hkpc, and is largest (by a factor of 3) for the post-mergers. This is
consistent with the results from \citep{ellison2013a} which show a
similar trend using the [OIII] luminosity of optical AGN indicative of
enhanced black hole accretion rate for close pairs and post-mergers
over isolated galaxies.}

\end{enumerate}

\section*{Acknowledgments} 

SS and WM gratefully acknowledge support from NASA grant NNX12AH53G for this project. SLE and DRP acknowledge the receipt of NSERC Discovery grants which
funded this research. RCH was partially supported by NASA through ADAP award NNX12AE38G and by the National Science Foundation through grant number 1211096. This paper benefited greatly from insightful discussions with Jessica Rosenberg and Nathan Secrest. We thank the referee for carefully reading the draft and providing comments that improved this paper.

This publication makes use of data products from the Wide-field
Infrared Survey Explorer, which is a joint project of the University
of California, Los Angeles, and the Jet Propulsion
Laboratory/California Institute of Technology, funded by the National
Aeronautics and Space Administration.

Funding for the SDSS and SDSS-II has been provided by the Alfred
P. Sloan Foundation, the Participating Institutions, the National
Science Foundation, the U.S. Department of Energy, the National
Aeronautics and Space Administration, the Japanese Monbukagakusho, the
Max Planck Society, and the Higher Education Funding Council for
England. The SDSS Web Site is http://www.sdss.org/.

The SDSS is managed by the Astrophysical Research Consortium for the
Participating Institutions. The Participating Institutions are the
American Museum of Natural History, Astrophysical Institute Potsdam,
University of Basel, University of Cambridge, Case Western Reserve
University, University of Chicago, Drexel University, Fermilab, the
Institute for Advanced Study, the Japan Participation Group, Johns
Hopkins University, the Joint Institute for Nuclear Astrophysics, the
Kavli Institute for Particle Astrophysics and Cosmology, the Korean
Scientist Group, the Chinese Academy of Sciences (LAMOST), Los Alamos
National Laboratory, the Max-Planck-Institute for Astronomy (MPIA),
the Max-Planck-Institute for Astrophysics (MPA), New Mexico State
University, Ohio State University, University of Pittsburgh,
University of Portsmouth, Princeton University, the United States
Naval Observatory, and the University of Washington.

\end{document}